\documentclass[twocolumn,prd,superscriptaddress,showpacs,floatfix,  
preprintnumbers, nofootinbib,hyperref]{revtex4} 
\usepackage{epsfig}   
\usepackage{bm}  
\usepackage{color}  
\usepackage{amsmath}

\newcommand{\ben}{\begin{equation}}
\newcommand{\een}{\end{equation}}

\newcommand{\gtrsim}{\,\rlap{\lower3.7pt\hbox{$\mathchar\sim$}}
\raise1pt\hbox{$>$}\,}
\newcommand{\lesssim}{\,\rlap{\lower3.7pt\hbox{$\mathchar\sim$}}
\raise1pt\hbox{$<$}\,}

\newcommand{\be}{\begin{equation}}
\newcommand{\ee}{\end{equation}}  
\newcommand{\bea}{\begin{eqnarray}}
\newcommand{\eea}{\end{eqnarray}}

\begin{document}

\title{Stability analysis of collective neutrino  oscillations in the supernova 
accretion phase   \\
with realistic energy and angle  distributions}

\author{Ninetta Saviano} 
\affiliation{II Institut f\"ur Theoretische Physik, Universit\"at Hamburg, Luruper Chaussee 149, 22761 Hamburg, Germany} 

\author{Sovan Chakraborty} 
\affiliation{II Institut f\"ur Theoretische Physik, Universit\"at Hamburg, Luruper Chaussee 149, 22761 Hamburg, Germany}
 
\author{Tobias Fischer} 
\affiliation{GSI, Helmholtzzentrum f\"ur Schwerionenforschung GmbH,
Planckstra{\ss}e 1
64291 Darmstadt, Germany}
\affiliation{Technische Universit\"at Darmstadt, Schlossgartenstra{\ss}e 9, 64289 Darmstadt,
Germany}

\author{Alessandro Mirizzi} 
\affiliation{II Institut f\"ur Theoretische Physik, Universit\"at Hamburg, Luruper Chaussee 149, 22761 Hamburg, Germany}

\begin{abstract}
We revisit our previous results on the matter
suppression of self-induced neutrino flavor conversions during a supernova (SN) accretion phase, 
performing a linearized stability analysis of the neutrino equations of motion, in the presence of
realistic
SN density profiles.
In our previous numerical study, we used a simplified  model based on an isotropic neutrino emission with
a single typical energy. Here,
 we take into account
  realistic    neutrino
energy and angle  distributions. We find that   multi-energy effects have a sub-leading impact
in the flavor stability of the SN neutrino fluxes with respect to our previous single-energy results. Conversely,  realistic forward-peaked neutrino angular
distributions would  enhance the matter suppression of the self-induced oscillations with respect
to  an isotropic neutrino emission. As a result, in our  models for iron-core SNe,  collective flavor  conversions
 have a  negligible impact on the characterization of the observable neutrino  signal
during the accretion phase. Instead, for a low-mass O-Ne-Mg core SN model, with lower
matter density profile and less forward-peaked angular distributions,
collective conversions   are  possible also at early times.
 
\end{abstract}


\maketitle

%%%%%%%%%%%%%%%%%%%%%%%%%%%%%%%%%%%%%%%%%%%
\section{Introduction} %%%%%%%%%%%%%%%%%%%%%%%%%%%%%%%%
%%%%%%%%%%%%%%%%%%%%%%%%%%%%%%%%%%%%%%%%%%%

Supernova (SN) neutrinos are important astrophysical messengers to probe
the flavor mixing in unique conditions~\cite{Raffelt:2012kt,Dasgupta:2010gr}. 
In particular,
renewed attention is being paid to collective features  of
flavor transformations%
~\cite{Fuller:2005ae,Duan:2005cp,Duan:2006an,Fogli:2007bk,Dasgupta:2007ws,Dasgupta:2009mg,Sawyer:2008zs,Friedland:2010sc,Dasgupta:2010ae,
Dasgupta:2010cd,
Duan:2010bf,Mirizzi:2010uz,Pehlivan:2011hp} induced by $\nu$-$\nu$ 
self-interactions~\cite{Pantaleone:1994ns,Qian:1993dg,Qian:1994wh}
in the deepest stellar regions, near the neutrino-sphere.
The observed collective phenomena
of synchronized~\cite{Pastor:2002we}  and pendular~\cite{Hannestad:2006nj} oscillations,
and the splits~\cite{Raffelt:2007cb,Raffelt:2007xt,Duan:2007bt} in the observable SN neutrino spectra 
are subjects of intense investigations (see~\cite{Duan:2010bg} for a recent review).
 
The implicit assumption in the characterization of the self-induced oscillations is related to the
flavor evolution in the deepest SN regions, being driven
by the only large neutrino densities $n_\nu$. However, during
the SN accretion phase (at post-bounce times $t_{\rm pb} \lesssim 0.5$~s) also the net electron density $n_e$ is
expected to be large. 
As 
pointed out in~\cite{EstebanPretel:2008ni}, when $n_e$ is not
negligible with respect to $n_\nu$, the large phase dispersion
induced by the matter for $\nu$'s traveling in different directions,
would partially or totally suppress the collective
oscillations through peculiar ``multi-angle effects''.
At this regard,
in two previous papers~\cite{Chakraborty:2011nf,Chakraborty:2011gd} we performed a numerical study of the matter
suppression during the accretion phase, characterizing  
the matter and the neutrino density profiles  with the  results from recent 
long-term SN hydrodynamical simulations~\cite{Fischer:2009af},
based on three flavor Boltzmann neutrino transport in spherical symmetry. 
Multi-angle numerical simulations of the non-linear neutrino equations of motion 
in the presence of large neutrino and matter densities    are computationally very demanding.
Indeed,
 these equations are ``stiff'' and their solutions generally involve a fast-changing combination of multi-frequency
oscillations. In order to reduce this numerical complexity, we 
assumed a simplified neutrino emission model. Namely, we considered all the neutrinos 
to be  emitted ``half-isotropically'' (i.e. with  all outward-moving angular modes 
equally occupied and all the backward-moving modes
empty) from a common neutrino-sphere with
a single representative energy.  In this framework, we found that the presence of a dominant matter
term inhibits the development of collective flavor conversions.
The matter suppression ranges from complete (when $n_e \gg n_\nu$)
to partial (when $n_e \gtrsim n_\nu$) suggesting,  in principle,  time-dependent features. 
In particular, for the iron-core SN models we analyzed, we found complete matter
suppression of the collective oscillations for post-bounce times $t_{\rm pb}\lesssim 0.2$~s and $t_{\rm pb}\gtrsim 0.4$~s and partial flavor conversions (with electron antineutrino survival 
probability $P_{ee}\simeq 0.5$) in the intermediate time range.

 Our works have stimulated further  independent 
investigations of these effects~\cite{Dasgupta:2011jf,Sarikas:2011am,Sarikas:2011jc}. 
In particular, in~\cite{Sarikas:2011am} it has been proposed to study the matter suppression of
the self-induced oscillations
applying the linearized stability analysis of the neutrino equations of motion, recently worked 
out in~\cite{Banerjee:2011fj}.
This method is particularly useful to circumvent the  challenges of full  numerical multi-angle simulations for the flavor 
evolution. These  are not necessary if one is  interested only in the issue of the flavor 
stability of the dense neutrino gas.
 Indeed,   the stability analysis would  allow one to determine  the possible onset of the flavor conversions, seeking  for
an exponentially growing solution of the eigenvalue problem, associated with the linearized equations of motion for
the neutrino ensemble. 
With this approach,
taking as benchmark for the SN density profiles and for the neutrino emission   a  
15.0~M$_\odot$ iron-core SN model from  the Garching group, 
the stability analysis has shown a complete suppression of the collective oscillations for all the duration
of the accretion phase~\cite{Sarikas:2011am}. Subsequently, the same technique has been applied to our 
numerical results~\cite{Sarikas:2011jc},
finding perfect agreement   with our sequence of complete and partial matter suppression.  

Motivated by these interesting papers, we find useful 
to improve our previous results 
applying  the stability analysis to our models, including also 
 realistic energy and angle neutrino distributions.
The plan of our work is as follows. 
In Sec.~2 we  introduce the SN models used to characterize the SN densities and, the neutrino  energy and angle 
fluxes.  
In Sec.~3 we  introduce the setup for the flavor-stability analysis, 
describing the non-linear equations for the neutrino flavor  evolution in SNe, and the consistency equations coming
from their linearization. 
In Sec.~4 we present our  results for the stability analysis of the matter suppression during the accretion phase
for two different SN progenitor masses. Finally, in Sec.~5 we comment on our results and we conclude.

%%%%%%%%%%%%%%%%%%%%%%%%%%%%%%%%%%%%%%%%%%%%%%%%%%%%%%%%%%%%%%%%%%%%%%%%%%%%%%%%%%%%
\section{Supernova models}
%%%%%%%%%%%%%%%%%%%%%%%%%%%%%%%%%%%%%%%%%%%%%%%%%%%%%%%%%%%%%%%%%%%%%%%%%%%%%%%%%%%%%

%%%%%%%%%%%%%%%%%%%%%%%%%%%%%%%%%%%%%%%%%%%
\begin{figure} 
\includegraphics[angle=0,width=1.\columnwidth]{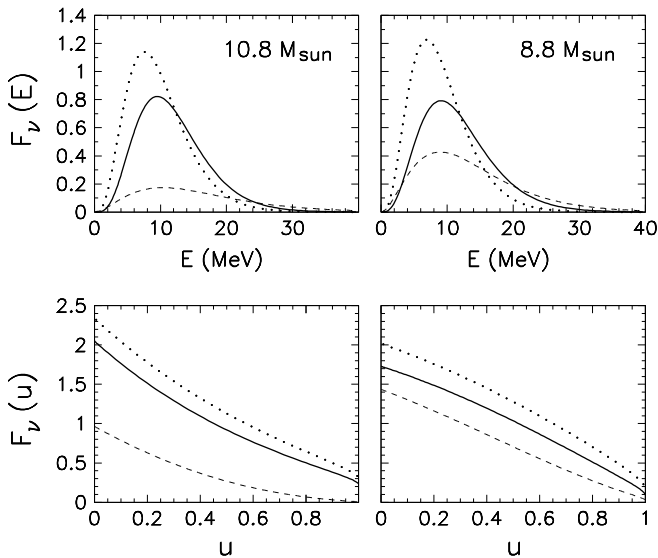}  
\vspace{0.2cm}
\caption{
Supernova neutrino flux spectra (in arbitrary units) for 
10.8~M$_\odot$ progenitor at $t_{\rm pb} = 0.225$~s (left panels) and 
for 8.8~M$_\odot$ progenitor at $t_{\rm pb} = 0.250$~s (right panels). 
The energy  (upper panels) and the angular (lower panels) spectra are shown 
for $\nu_e$ (dotted curves), ${\overline\nu}_e$ (continuous curves)
and $\nu_x$ (dashed curves) (see the text for details).
\label{fig1}}
\end{figure}
%%%%%%%%%%%%%%%%%%%%%%%%%%%%%%%%%%%%%%%%%%%

We considered the core-collapse supernova simulations
of massive progenitor stars with 8.8 and 10.8~M$_\odot$ progenitor 
 from Ref.~\cite{Fischer:2009af}, taken as benchmark
for our numerical study  in~\cite{Chakraborty:2011gd}.
The first SN  belongs to the class of O-Ne-Mg-core progenitors
\cite{Nomoto:1983,Nomoto:1987}
and represents the threshold between thermonuclear
explosions and core-collapse supernovae \cite{Kitaura:2006, Fischer:2009af}.
The second is an iron-core progenitor \cite{Woosley:2002}.
All models were evolved consistently through core collapse,
bounce and the early post-bounce phase up to several seconds
after the onset of explosion~\cite{Fischer:2009af}.
We remind that in order to trigger the explosion for the 10.8~M$_\odot$ progenitor model, 
the heating rates have been artificially enhanced in the gain region where neutrinos 
deposit energy in order to revive the stalled bounce shock. 

In our previous study, we schematically took
 a mono-energetic $\nu$ ensemble,  with a representative
energy  $E=15$~MeV.  Neutrinos
of different species were assumed to be
emitted half-isotropically  by a common spherical ``neutrino-sphere,''
in analogy with a blackbody emission. However,
realistic supernova simulations show
that $\nu$  angular distributions at the decoupling are far from being half-isotropic and are flavor-dependent
(see, e.g.,~\cite{Sarikas:2011am,Ott:2008jb}).
In order to fix a common neutrino-sphere radius $r=R$ for the flavor evolution, consistently with our 
choice in~\cite{Chakraborty:2011gd}  we take
the radius at which the
$\nu_e$'s angular distribution has no longer significant backward flux, i.e. a few $\%$ of
the total one. This typically is in the range $R\sim 50-100$~km (see Fig.~4 in~\cite{Chakraborty:2011gd}). 
It is  convenient to parameterize every angular mode in terms of its emission angle $\theta_R$ relative to the radial direction
of the neutrino-sphere.  For
a half-isotropic distribution the occupation numbers are distributed as $d n/d \cos \theta_R = \textrm{const}$,
or equivalently the radial fluxes are distributed as $d F/ d \cos \theta_R \propto \cos\theta_R$~\cite{EstebanPretel:2007ec}.
A further simplification is obtained if one labels the different angular modes in terms of the variable $u=\sin^2\theta_R$,
as in~\cite{EstebanPretel:2007ec,Banerjee:2011fj}. Note that 
for a half-isotropic emission at the neutrino-sphere  the $\nu$ angular distribution of the radial fluxes
 is a box spectrum in $0\leq u \leq 1$, since $ d \cos \theta_R/du\propto (\cos\theta_R)^{-1}$  cancels the $\cos\theta_R$
 dependence previously mentioned.
 
 From SN simulations of~\cite{Fischer:2009af}  we extract the angle and energy distributions $F_{\nu_{\alpha}}(E,u)$ of the different neutrino species. 
 In Fig.~1 we show the
(angle-integrated)  flux energy spectra $F_{\nu_{\alpha}}(E)$ (upper panels) and the 
(energy-integrated) flux angular spectra $F_{\nu_{\alpha}}(u)$ (lower panels)
 for $\nu_e$, ${\overline\nu}_e$,
$\nu_x$, where this latter indicates the non-electron flavors. Here, we are showing these fluxes for  the  10.8~M$_\odot$ SN progenitor
at $t_{\rm pb}=225$~ms (left panels), and for the  8.8~M$_\odot$ SN progenitor
at $t_{\rm pb}=250$~ms (right panels).
The angular variable $0\leq u \leq 1$ is based on $R=69$~km for the 10.8~M$_\odot$ model, and on 
$R= 47$~km for the 8.8~M$_\odot$ model.
The energy and angular distributions are normalized to the total neutrino number fluxes of the different species (in arbitrary units in the Figure). 

Concerning the neutrino energy distributions for the different flavors, 
the angle-integrated energy spectra are well-represented
by the renowned  form~\cite{Keil:2002in,Raffelt:2003en} 
%%%%%%%%%%%%%%%%%%%%%%%%%%%%%%%%%%%%%%%%%%%%%%%%%%%%
\begin{equation}
F_{\nu_{\alpha}}(E) = \frac{L_\nu}{{\langle E_\nu\rangle}^2}\frac{(1+\beta)^{1+\beta}}{\Gamma(1+\beta)}\left(\frac{E}{\langle E_\nu\rangle}\right)^\beta\exp\left[-(1+\beta)\frac{E}{\langle
E_\nu\rangle}\right]\, ,
 \label{eq:varphi}
\end{equation}
%%%%%%%%%%%%%%%%%%%%%%%%%%%%%%%%%%%%%%%%%%%%%%%%%%%%%%%%%%%%%%%%%%%%%%%%%%
where for a given flavor
$L_{\nu}$ is the neutrino luminosity, $\langle E_\nu\rangle$ is the average energy and
 $\beta$ is  the energy-shape parameter, defined as~\cite{Keil:2002in,Raffelt:2003en}
\begin{equation}
\beta=\frac{2\langle E_\nu \rangle^2-\langle E_\nu^2\rangle}{\langle E_\nu^2\rangle-
\langle E_\nu\rangle^2} \,, \label{alphadef}
\end{equation}
i.e. it is a dimensionless parameter containing information on the second moment of the distribution, $\langle E_\nu^2\rangle$. 
 In general,  $L_\nu$, $\langle E_\nu\rangle$ and $\langle E_\nu^2\rangle$ are all functions of time, and are  extracted directly from the simulations at hand (see Fig.~2 in~\cite{Fischer:2009af}).
 In the case of 10.8~M$_\odot$  we find a typical total number flux hierarchy expected
during the accretion phase, numerically
 $F_{\nu_e} : F_{{\overline\nu}_e}:
 F_{\nu_x} = 1.20:1.00:0.34$. The    first part of the hierarchy
is caused by the deleptonization of the collapsed core
whereas the second is caused by the absence of charged current
interactions for neutrino species other than $\nu_e$ and 
${\overline\nu}_e$. 
Passing now to the case of the low-mass 8.8~M$_\odot$ SN progenitor,
we realize that 
the spectrum of the non-electron flavors is less suppressed with respect
to the electron (anti)neutrino spectra, 
namely  $F_{\nu_e} : F_{{\overline\nu}_e}:
 F_{\nu_x} = 1.23:1.00:0.72$.
This different behavior is due to the fact that in low-mass core SNe, the duration of the 
accretion phase is very short (i.e. $t_{\rm pb} \lesssim 0.03$~s
in Fig.~2 of Ref.~\cite{Fischer:2009af}). At later times,  the neutrino luminosities can be approximated by diffusion
and the spectra are determined from thermalization processes (e.g., inelastic
scattering on electrons/positrons and elastic scattering on nucleons). 
Since these neutral-current processes do not distinguish  the different flavors, this
leads to reduced spectral differences between electron and non-electron species 
(see, e.g.,~\cite{Fischer:2011cy}).

Considering now to the angular flux spectra, we realize that in the case of  the 10.8~M$_\odot$ model, these 
are  significantly forward enhanced (i.e. peaked at small $u$), with respect to the
half-isotropic emission model.
Remarkably, the angular distributions of the three $\nu$ species  at the conventional neutrino-sphere 
are rather different: since the $\nu_x$'s  decouple at smaller radii with respect to $\nu_e$'s and 
$\overline\nu_e$'s, their 
distributions are more suppressed in the direction tangential to the neutrino-sphere
(i.e. at $u\sim 1$).
In the case of 8.8~M$_\odot$ the angular spectra for the electron 
species are less forward-enhanced than in the previous case, 
and with less pronounced differences with  the non-electron  species. 
As for the energy spectra, this  behavior 
is due to the dominant role of the neutral-current processes in the spectra formation.  
We also note that for the models we considered the angular distributions do not cross at any $u$.~\footnote{As 
recently discussed in~\cite{Mirizzi:2011tu}, possible crossings in the angular distributions of different flavors could lead to an enhancement of the
flavor instability.}

%%%%%%%%%%%%%%%%%%%%%%%%%%%%%%%%%%%%%%%%%%%%%%%%%%%%%%%%%%%%%%%%%%%%%%%%%%%%%%%%%%%%
\section{Setup of the stability analysis}
%%%%%%%%%%%%%%%%%%%%%%%%%%%%%%%%%%%%%%%%%%%%%%%%%%%%%%%%%%%%%%%%%%%%%%%%%%%%%%%%%%%%%

%%%%%%%%%%%%%%%%%%%%%%%%%%%%%%%%%%%%%%%%%%%%%%%%%%%%%%%%%%%%%%%%%%%%%%%%%%%%%%%%%%%%
\subsection{Equations of motion}
%%%%%%%%%%%%%%%%%%%%%%%%%%%%%%%%%%%%%%%%%%%%%%%%%%%%%%%%%%%%%%%%%%%%%%%%%%%%%%%%%%%%%

We work in  a two-flavor oscillation scenario, associated to the 
atmospheric mass-square difference $\Delta m^2_{\rm atm}= 2 \times 10^{-3}$~eV$^2$ and
and with the small (matter suppressed) in-medium mixing $\Theta_{\rm eff} = 10^{-3}$~\cite{Kuo:1989qe}.  
We will always assume inverted mass hierarchy ($\Delta m^2_{\rm atm}<0$) where $\nu$'s can
exhibit flavor instabilities for the flux ordering present during the accretion phase~\cite{Fogli:2007bk}. 
For this flux hierarchy, three-flavor effects, associated with the solar mass splitting are negligible~\cite{Dasgupta:2007ws}. 
Following~\cite{Banerjee:2011fj}, we write the equations of motion for the flux matrices  $\Phi_{E,u}$  as function of the radial coordinate.
The diagonal $\Phi_{E,u}$ elements are
the ordinary number fluxes $F_{\nu_{\alpha}}(E,u)$ 
integrated
over a sphere of radius $r$.
We normalize the flux matrices to the total ${\overline\nu}_e$ number flux $n_{{\bar\nu}_e}$ at the neutrino-sphere.
 Conventionally, we use negative $E$ and negative
number 
fluxes for anti-neutrinos. The off-diagonal elements,
which are initially zero, carry a phase information due to 
flavor mixing.
Then, the equations of motion read~\cite{Banerjee:2011fj,Sigl:1992fn}
%.....................................................
\begin{equation}
\textrm{i}\partial_r \Phi_{E,u}=[H_{E,u},\Phi_{E,u}] \,\ 
\label{eq:eom1}
\end{equation}
%.........................................................
with the Hamiltonian~\cite{Pantaleone:1992eq,Qian:1994wh,Banerjee:2011fj,Sigl:1992fn}
%.......................................................
 \begin{eqnarray}
& & H_{E,u} = \frac{1}{v_{u}} \left(\frac{M^2}{2E} + \sqrt{2} G_F N_l \right) \nonumber \\
 &+& \frac{\sqrt{2}G_F}{4\pi r^2}\int_{-\infty}^{+\infty}d E^\prime
 \int_{0}^{1}du^\prime \left(\frac{1-v_{u}v_{u^\prime}}{v_{u}v_{u^\prime}}
 \right)\Phi_{E^\prime,u^\prime} \,\ .
 \label{eq:eom2}
 \end{eqnarray}
%......................................................... 
 The matrix $M^2$ of neutrino mass-squares causes vacuum
flavor oscillations. 
The matrix $N_l= \textrm{diag}(n_e,0,0)$ in flavor basis, contains the net electron density and 
is responsible for the Mikheyev-Smirnov-Wolfenstein (MSW) matter effect~\cite{Matt}   with the ordinary background. 
Finally, 
 the term at second line represents the $\nu$-$\nu$ refractive 
term. 
In particular, the 
 factor proportional to the neutrino velocity 
$v_{u} = (1-u R^2/r^2)^{1/2}$~\cite{EstebanPretel:2007ec} in 
the $\nu$-$\nu$ interaction term
implies ``multi-angle'' effects for neutrinos moving on different trajectories~\cite{Pantaleone:1992eq,Qian:1994wh, Duan:2006an,Sigl:1992fn}. In order to properly simulate numerically this effect one needs 
to  follow a large number
$[{\mathcal O}(10^3)]$ of interacting neutrino modes.

%%%%%%%%%%%%%%%%%%%%%%%%%%%%%%%%%%%%%%%%%%%%%%%%%%%%%%%%%%%%%%%%%%%%%%%%%%%

%%%%%%%%%%%%%%%%%%%%%%%%%%%%%%%%%%%%%%%%%%%%%%%%%%%%%%%%%%%%%%%%%%%%%%%%%%%%%%%%
\subsection{Stability conditions}
%%%%%%%%%%%%%%%%%%%%%%%%%%%%%%%%%%%%%%%%%%%%%%%%%%%%%%%%%%%%%%%%%%%%%%%%%%%%%
In order to perform the stability analysis we closely follow the
prescriptions presented in~\cite{Banerjee:2011fj} and summarized in the following. 
At first we switch to the 
frequency variable $\omega= \Delta m^2_{\rm atm}/2E$ so that $E(\omega)=|\Delta m^2_{\rm atm}/2\omega|$ and  we introduce the 
neutrino flux difference distributions $g_{\omega,u}\equiv g(\omega,u)$ defined as
\begin{eqnarray}
&&g_{\omega,u}=\frac{|\Delta m_{\rm atm}^2|}{2\omega^2}\times\bigg\{\Theta(\omega)\left[F_{\nu_e}(E(\omega),u)-F_{\nu_x}(E(\omega),u\right)]\nonumber\\
&&+\Theta(-\omega)\left[F_{\nu_x}(E(\omega),u)-F_{\overline\nu_e}(E(\omega),u)\right]\bigg\}
\end{eqnarray}
normalized  to the total ${\overline\nu}_e$ flux at the neutrino-sphere. 
Then, we   write the flux matrices in the form~\cite{Banerjee:2011fj}
%...........................................................
\begin{equation}
\Phi_{\omega,u}= \frac{\textrm{Tr}\Phi_{\omega,u}}{2}+
\frac{g_{\omega,u}}{2}
\left( \begin{array}{cc} s_{\omega,u} &  S_{\omega,u} \\
S^{\ast}_{\omega,u} & -s_{\omega,u} 
\end{array} \right) \,\ ,
\end{equation}
%..............................................................
where $\textrm{Tr}\Phi_{\omega,u}$ is conserved and then irrelevant for the flavor conversions, and the initial conditions for the  
``swapping matrix'' in the second term on the right-hand side are $s_{\omega,u}=1$ and $S_{\omega,u}=0$.
Self-induced flavor transitions  start when the off-diagonal term $S _{\omega,u}$ grows
 exponentially.

In the small-amplitude limit $|S _{\omega,u}|\ll 1$, 
and at far distances from the neutrino-sphere $r \gg R$,
the linearized evolution equations for  $S _{\omega,u}$
in inverted mass hierarchy ($\Delta m^2_{\rm atm}<0$)
assume the form~\cite{Banerjee:2011fj}
%..................................................
\begin{eqnarray}
\textrm{i}\partial_r S _{\omega,u} &=& [\omega + u(\lambda +\epsilon \mu)] S _{\omega,u} \nonumber \\
&-&\mu \int du^\prime d\omega^\prime (u+u^\prime)g_{\omega^\prime, u^\prime} S _{\omega^\prime,u^\prime} \,\ ,
\label{eq:stab}
\end{eqnarray}
%...........................................................
where 
%.........................................................
\begin{equation}
\epsilon = \int du \,\ d\omega \,\ g_{\omega,u} \,\ ,
\label{eq:asy}
\end{equation}
%........................................................
quantifies the ``asymmetry'' of the neutrino spectrum, normalized to the total ${\overline\nu}_e$ 
number flux.
The $\nu$-$\nu$ interaction strength is given by
%.................................................................
\begin{eqnarray}
\mu &=& \frac{\sqrt{2}G_F n_{{\bar\nu}_e}(R)}{4 \pi r^2}\frac{R^2}{2 r^2} \nonumber \\
&=& \frac{3.5 \times 10^{9}}{r^4} \left(\frac{L_{{\overline\nu}_e}}{10^{52} \,\ \textrm{erg/s}}\right)
\left(\frac{15 \,\ \textrm{MeV}}{\langle E_{{\overline\nu}_e} \rangle}\right) 
 \left(\frac{R}{10 \,\ \textrm{km}} \right)^2 ,  \nonumber
\end{eqnarray}
%..................................................................
while  ordinary matter background term is given by 
%............................................................
\begin{eqnarray}
\lambda &=& \sqrt{2} G_F n_e \frac{R^2}{2 r^2} \nonumber \\
&=& \frac{0.95 \times 10^8 }{r^2} \left(\frac{Y_e}{0.5} \right)
\left(\frac{\rho}{10^{10} \textrm{g}/\textrm{cm}^3} \right) 
\left(\frac{R}{10 \,\ \textrm{km}} \right)^2 , \nonumber 
\end{eqnarray}
%.............................................................
where $Y_e$ is the net electron fraction,  and $\rho$ is the matter density. 
The radial distance $r$ is expressed in km, 
while the numerical values of $\mu$ and  $\lambda$ in the two previous equations 
are quoted in km$^{-1}$, as appropriate for the SN case. 

One can write the solution of the linear differential equation [Eq.~(\ref{eq:stab})]
in the form $S _{\omega,u} = Q _{\omega,u} e^{-i\Omega r}$ with complex frequency
$\Omega= \gamma + i \kappa$ and eigenvector $Q _{\omega,u}$. A solution with 
$\kappa >0$ would indicate an exponential increasing $S _{\omega,u}$, i.e. an instability.
The solution of Eq.~(\ref{eq:stab}) can then be recast in the form of an eigenvalue equation
for $Q _{\omega,u}$. Splitting this equation into its real and imaginary parts one arrives
at two real equations that have to be satisfied~\cite{Banerjee:2011fj}
%.......................................................................
\begin{eqnarray}
(J_1-\mu^{-1})^2 &=& K_1^2 + J_0 J_2 -K_0 K_2 \,\ , \nonumber \\
(J_1-\mu^{-1}) &=& \frac{J_0 K_2 + K_0 J_2}{2 K_1} \,\ ,
\label{eq:JnKn}
\end{eqnarray}
%..........................................................
where
%...............................................
\begin{eqnarray}
J_n &=& \int d\omega du \,\ g_{\omega,u} u^n \frac{\omega +u (\lambda +\epsilon \mu) -\gamma}
{[\omega +u(\lambda +\epsilon \mu) -\gamma]^2 + \kappa^2} \,\ , \nonumber \\
K_n &=& \int d\omega du \,\ g_{\omega,u} u^n \frac{\kappa}
{[\omega +u(\lambda +\epsilon \mu)-\gamma]^2 + \kappa^2} .
\label{eq:consit}
\end{eqnarray}
%.................................................
A flavor instability is present whenever Eqs.~(\ref{eq:JnKn}) admit a  solution $(\gamma, \kappa)$. 

%%%%%%%%%%%%%%%%%%%%%%%%%%%%%%%%%%%%%%%%%%%%%%%%%%%%%%%%%%%%%%%%%%%%%%%%%%%%%%%%%%%%
\section{Application to our supernova models}
%%%%%%%%%%%%%%%%%%%%%%%%%%%%%%%%%%%%%%%%%%%%%%%%%%%%%%%%%%%%%%%%%%%%%%%%%%%%%%%%%%%%%

In this Section we will present our results for the stability analysis 
of the self-induced flavor conversions for two representative SN simulations based on different
progenitor masses. 

\subsection{10.8 M$_{\odot}$ progenitor mass}

We start our investigation with the case of the 10.8~M$_{\odot}$ iron-core supernova. 
We briefly remind the reader our previous results presented in Ref.~\cite{Chakraborty:2011gd}.
For this SN model, 
the net electron density $n_e$ and the neutrino densities $n_\nu$ for different post-bounce times
were shown in Fig.~5 of our previous paper~\cite{Chakraborty:2011gd}. 
The numerical results of the multi-angle flavor  evolution for our schematic 
model with neutrinos of different species all emitted half-isotropically with a single energy, were shown 
in Fig.~7.
In particular, it was represented the survival probability of electron antineutrinos $P_{ee}$, at different
post-bounce times in the presence of matter effects and for $n_e=0$. 
 The matter effects produced a  complete 
suppression of the flavor conversions for $t_{\rm pb}\lesssim 0.2$~s and $t_{\rm pb}\gtrsim 0.4$~s
(where $n_e/n_{{\bar\nu}_e} \gg 1$, see Fig.~8 in~\cite{Chakraborty:2011gd}) and partial flavor conversions (with electron-survival 
probability $P_{ee}\simeq 0.5$) at the intermediate times (where $n_e/n_{{\bar\nu}_e} \gtrsim 1$).

We now  compare these numerical results with  what we find  using the stability analysis.
For the
 same post-bounce times 
considered in our previous paper,  we show in Fig.~2 the radial evolution of  the eigenvalue $\kappa$ determined 
from the solution of Eqs.~(\ref{eq:JnKn}).
We consider the following cases: \emph{ (a)}   $n_e=0$ and a half-isotropic neutrino emission (dashed curves),
\emph{ (b)}
 dense matter effects and
   a half-isotropic neutrino emission
(continuous curves) and,  \emph{ (c)} dense  matter effects and non-trivial  neutrino angular distributions 
(dotted curves). In all these cases we have always assumed quasi-thermal neutrino  energy
distributions, parametrized as in Eq.~(\ref{eq:varphi}).

%%%%%%%%%%%%%%%%%%%%%%%%%%%%%%%%%%%%%%%%%%%%
\begin{figure}[!t] 
\includegraphics[angle=0,width=1.\columnwidth]{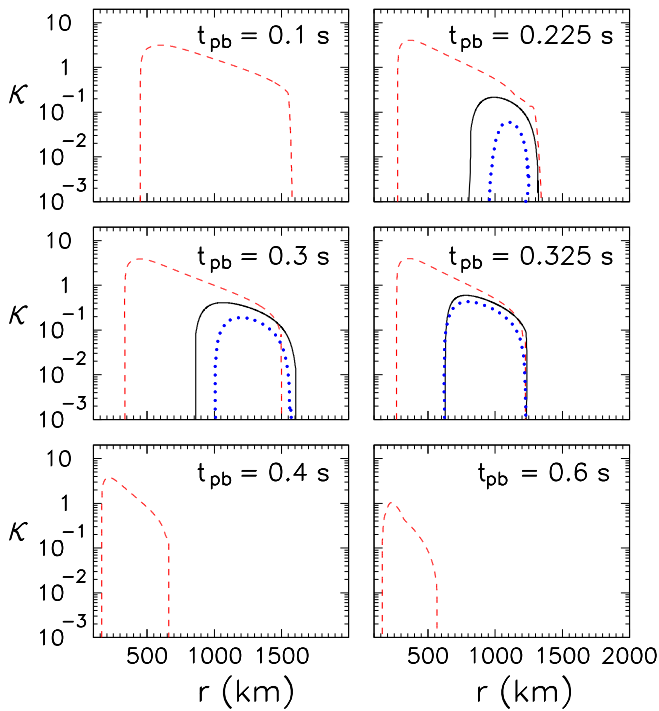}  
\caption{
10.8 M$_{\odot}$ progenitor mass. Radial evolution of the 
$\kappa$ function at different post-bounce times
 with $n_e=0$ for a half-isotropic neutrino emission
(dashed curves)  and in presence of matter effects, with 
a half-isotropic neutrino emission (continuous curves) and 
with flavor-dependent angular distributions (dotted curves).
Quasi-thermal $\nu$ energy spectra are assumed (see the text for  details). 
\label{fig2}}
\end{figure}
%%%%%%%%%%%%%%%%%%%%%%%%%%%%%%%%%%%%%%%%%%% 

We start discussing our results for the case of $n_e=0$. 
We realize that when the neutrino system enters into an unstable regime ($\kappa >0$), the $\kappa$
function rapidly grows from zero to a peak value greater than one.  
Comparing this result with the numerical solution presented 
in  Fig.~7 of~\cite{Chakraborty:2011gd}, we find perfect agreement between the onset of the self-induced
flavor conversions determined numerically and the position of the  peak in the $\kappa$ function. 
Remarkably, the multi-energy effects included in the stability analysis do not seem to
produce major changes in the onset of the flavor conversions with respect to what
observed in our previous  mono-energetic investigation. 
Of course, once  conversions are triggered, multi-energy effects would play a crucial 
role in determining the splitting features in the final neutrino spectra. 
For their characterization, multi-energy and multi-angle simulations of the flavor 
evolution  are mandatory (see, e.g.,~\cite{Mirizzi:2010uz}). 

We  now discuss the case with realistic matter density profiles and 
a half-isotropic neutrino emission. 
As expected, the flavor instability is strongly suppressed with respect to the previous
case with $n_e=0$. 
In particular, the $\kappa$ function is non-zero  only at intermediate post-bounce
times, i.e. $t_{\rm pb}=0.225, 0.3, 0.325$~s in Fig.~2, consistently with the occurrence
of the flavor conversions in~\cite{Chakraborty:2011gd}.  
In these cases the rise of the $\kappa$ function is  shifted at larger radii (by $\sim 400-500$~km) 
with respect to the
case with $n_e=0$. In particular, the instability occurs when $n_e \sim n_{{\bar\nu}_e}$. 
 Moreover, the peak value of $\kappa$ in these cases
 reaches at most $\sim 0.5$,
implying a slower growth of the instability with respect to the  case with $n_e=0$.
In the intermediate time snapshots, where the flavor conversions are not completely  suppressed,  we find an agreement between the onset 
of Fig.~7 in~\cite{Chakraborty:2011gd} and the position of the  peak in the $\kappa$ function. 

Finally, we  consider the case in which also the flavor-dependent forward-peaked neutrino angular
distributions are taken into account. 
We find that the $\kappa$ function is further suppressed with respect
to the half-isotropic case. 
This is consistent  with the expectation that the $\nu$-$\nu$ strength is 
weaker for forward-peaked distributions, making the system more stable under the effect
of the matter.
We also note that at $t_{\rm pb}=0.325$~s when the accretion is at the  end,  the effect
of the angular-distributions is less pronounced, since these become less forward-enhanced,
as discussed in Sec.~2.  

In  order to validate these findings  we also performed multi-angle and multi-energy 
simulations of the flavor evolution, for the cases with realistic matter densities and neutrino
 energy and angle distributions. We used
$N_u=1200$ angular modes and $N_E=80$ energy modes.
Our main goal is  to determine if flavor conversions are present or not, without being  interested in the exact final
outcome of the numerical simulations. Therefore,  we did not require the  perfect numerical convergence
of our results. 
In the cases in which we have not found a positive $\kappa$ with the stability analysis,
also the numerical evolution has shown no flavor conversion. 
Moreover at $t_{\rm pb}=0.225$~s, where $\kappa <0.1$,
 the numerical simulation gives a  complete suppression of the flavor conversions.
 Conversely,  for the two other
 post-bounce times ($t_{\rm pb}=0.3, 0.325$~s)  where $\kappa$ reaches a peak between 0.2 and 0.4, we find partial flavor 
conversions. 
However, in runs with different numbers of angular and energy 
modes  the final value of $P_{ee}$ ranges between $\sim 0.8-0.9$,
suggesting that these partial flavor conversions would be practically negligible in the characterization
of the supernova neutrino signal during the accretion phase.
 
Finally, we mention that we applied the  stability analysis also to   the case of a 
18.0~M$_{\odot}$ iron-core supernova, considered in~\cite{Chakraborty:2011gd}.
However,  since the results obtained are similar to the ones
discussed in this Section, for the sake of the brevity, we do not show them here.

\subsection{8.8 M$_{\odot}$ progenitor mass}

We  now  analyze the case of a SN
with a low-mass 8.8 M$_{\odot}$  O-Ne-Mg core. 
We remind that in this case there is an absence of an extended accretion phase, since the explosion
succeeds very shortly after the core-bounce.
For this model, 
the net electron density $n_e$ and the neutrino densities $n_\nu$ at different post-bounce
times ($t_{\rm pb}\leq 0.25$~s) 
are shown in Fig.~11 of~\cite{Chakraborty:2011gd}.
 The numerical results of the multi-angle flavor  evolution for our schematic 
model with neutrinos emitted half-isotropically with a single energy were shown in Fig.~13
 of~\cite{Chakraborty:2011gd}.
From this Figure one can realize that the matter suppression of self-induced flavor conversions 
is never complete, since the matter density is very low with respect to iron-core progenitor.
In Fig.~3 we plot the corresponding radial evolution of the $\kappa$ 
function, for the  \emph{ (a)},\emph{ (b)},\emph{ (c)}
 cases introduced in the previous Section.
In the case of a half-isotropic neutrino emission with  $n_e=0$
(dashed curves) we find once more a perfect agreement between the   peak of 
the $\kappa$ function and the onset of the flavor transitions found numerically. 

We now discuss  the matter effects. At first,  
we realize that the presence of non-trivial
angular distributions (dotted curves) does not lead to a  further suppression of the instability
with respect to the case with a half-isotropic emission (continuous curves).
This different behavior with respect to what we have seen in the case of
  10.8 M$_{\odot}$ SN model,
 is consistent with what is shown in Fig.~1, i.e. the angular spectra of different flavors
for the
 8.8 M$_{\odot}$ SN are significantly
 less forward-peaked than in the case of the  10.8 M$_{\odot}$ SN.
 Therefore, their effect on the flavor stability is less pronounced. 
Looking in details at the different time snapshots in Fig.~3, we realize that at 
$t_{\rm pb}=0.08$~s (when $n_{{\bar\nu}_e} \lesssim n_e$)  the $\kappa$ function presents a
peculiar shape coming from the interplay between the self-induced and the matter effects
with a comparable strength. 
The $\kappa$ curve broadens at smaller $r$ with respect to the case with $n_e=0$. 
The peak of $\kappa$ is slightly reduced by the matter effects.
 However, its position remains the same, giving the onset of the flavor conversions
 found  numerically. 
At later times, when $n_e$ dominates  over $n_{{\bar\nu}_e}$, the matter suppression becomes more 
relevant.
However, 
since at most $n_e \gtrsim 2 n_{{\bar\nu}_e}$, 
the suppression is never as  strong as in the case of the iron-core SNe.  Also in this case, the matter effect shifts 
at larger radii the onset of the flavor conversions.

%%%%%%%%%%%%%%%%%%%%%%%%%%%%%%%%%%%%%%%%%%%
\begin{figure} 
\includegraphics[angle=0,width=1.\columnwidth]{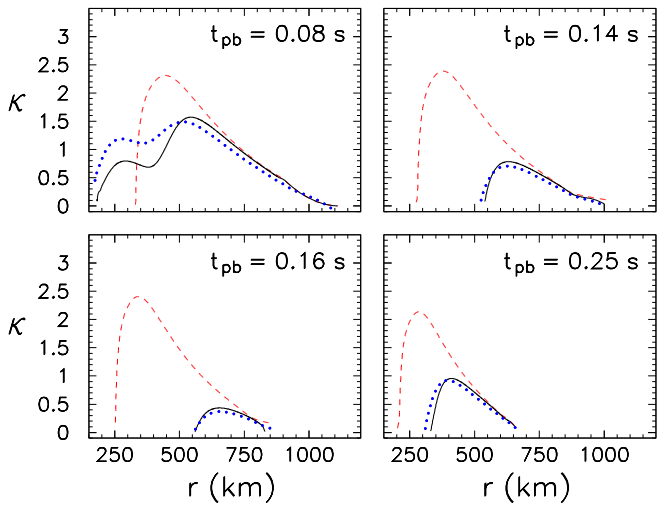}  
\caption{
8.8 M$_{\odot}$ progenitor mass. Radial evolution of the 
$\kappa$ function at different post-bounce times
 with $n_e=0$ for a half-isotropic neutrino emission
(dashed curves)  and in presence of matter effects, with 
a half-isotropic neutrino emission (continuous curves) and 
with flavor-dependent angular distributions (dotted curves).
Quasi-thermal $\nu$ energy spectra are assumed (see the text for details). 
\label{fig3}}
\end{figure}
%%%%%%%%%%%%%%%%%%%%%%%%%%%%%%%%%%%%%%%%%%% 

%%%%%%%%%%%%%%%%%%%%%%%%%%%%%%%%%%%%%%%%%%%%%%%%%%%%%%%%%%%%%%%%%%%%%%%%%%%%%%%%%%%%%%
\section{Conclusions}
%%%%%%%%%%%%%%%%%%%%%%%%%%%%%%%%%%%%%%%%%%%%%%%%%%%%%%%%%%%%%%%%%%%%%%%%%%%%%%%%%%%%%%%%%%

We have performed a linear stability analysis of the self-induced flavor conversions during the 
accretion phase for two SN models with different progenitor masses. We characterize  the SN densities,
 the neutrino energy and angular spectra, with results from recent SN hydrodynamical simulations.
We compared this method with the numerical results of the flavor evolution  presented in our previous
 paper~\cite{Chakraborty:2011gd}, where
we assumed a schematic $\nu$ emission, with that all the neutrinos  streaming half-isotropically
from the neutrino-sphere with a single representative energy. 
For the case of an iron-core 10.8 M$_{\odot}$  SN (and 18.0 M$_{\odot}$, not shown) we found that
the continuous $\nu$ energy spectra do not play a crucial role in the issue of the 
stability of the neutrino ensemble with respect to our previous single-energy results.
Conversely,  the presence of forward-peaked $\nu$ angular distributions significantly
reduces the strength of the $\nu$-$\nu$ interaction term. This effect would enhance the suppression of the
self-induced oscillations, due to  the dense matter 
term.
We find that the flavor instability is  completely suppressed for large part of the duration of the 
accretion phase, excepct for a small time window around $t_{\rm pb}\sim 0.3$~s. However, we checked with 
multi-energy and multi-angle simulations that the effect of  these partial  flavor conversions 
would be negligible in the characterization of the observable supernova neutrino flux.
Our result is in agreement with  the stability analysis recently performed 
with a  15.0 M$_{\odot}$ SN model from the Garching group~\cite{Sarikas:2011am}. In that paper, a complete matter suppression 
of the self-induced flavor conversions was found for all the duration of the accretion phase. 
In the case of  a low-mass O-Ne-Mg SN with 8.8 M$_{\odot}$ progenitor, where  the
 accretion phase is extremely short, 
the matter density profile is lower  and the $\nu$ angular distributions less forward-peaked 
than in iron-core models. As a consequence, 
we found that also with realistic angular distributions flavor conversions   would be    possible at early times, i.e. at $t_{\rm pb}\lesssim 0.2$~s.
The different pattern of the self-induced flavor conversions for these two SN progenitors could be, 
at least in principle, a tool to distinguish between iron-core and O-Ne-Mg core SNe.

The matter suppression of the self-induced flavor conversions for iron-core supernovae would imply
that the original neutrino spectra will be processed only by the ordinary 
 Mikheyev-Smirnov-Wolfenstein  effect in the outer stellar layers.
This effect would allow to distinguish the neutrino mass hierarchy through the  
Earth matter effects~\cite{Dighe:2003jg} or the rise time of the SN 
neutrino signal~\cite{Chakraborty:2011ir}, in the case of  
a $\theta_{13}$ neutrino mixing angle as ``large''
as currently measured by the Daya Bay~\cite{An:2012eh} and Reno~\cite{Ahn:2012nd}  reactor experiments. These recent measurement confirm
and greatly strengthen the significance of early hints suggested by the long-baseline $\nu_{\mu}$-$\nu_e$ 
experiments~\cite{Abe:2011sj,Adamson:2011qu} and
Double Chooz reactor experiment~\cite{Abe:2011fz}, especially when analysed in 
combination with other oscillation data~\cite{Fogli:2011qn}.
This intriguing 
consequence implies that the characterization of the $\nu$ flavor
evolution during the accretion phase is crucial  for the interpretation of the observable SN $\nu$ signal.
At this regard in~\cite{Dasgupta:2011jf}, due to a different choice of the neutrino angular distributions, 
a less pronounced matter suppression  has been found for iron-core SNe.
In future it will be mandatory to apply the stability analysis also to other SN models 
in order to understand how generic are the results obtained till now. 

%%%%%%%%%%%%%%%%%%%%%%%%%%%%%%%%%%%%%%%%%%%%%%%%%%%%%%%%%%%%%%%%%%%%%%
\section*{Acknowledgements} %%%%%%%%%%%%%%%%%%%%%%%%%%%%%%%%%%%%%%%%%%%%%%%%
%%%%%%%%%%%%%%%%%%%%%%%%%%%%%%%%%%%%%%%%%%%%%%%%%%%%%%%%%%%%%%%%%%%%%%

We thank P.D.~Serpico for interesting discussions during the development of this project.
We also acknowledge G.~Raffelt and G.~Sigl for reading the manuscript and for useful comments on it.
The work of S.C.,  A.M., N.S.  was supported by the German Science Foundation (DFG)
within the Collaborative Research Center 676 ``Particles, Strings and the
Early Universe''.
T.F. acknowledges support from the Swiss National Science Foundation (SNF)
 under grant~no.~PBBSP2-133378 and HIC for FAIR.

%%%%%%%%%%%%%%%%%%%%%%%%%%%%%%%%%%%%%%%%%%%
\section*{References} %%%%%%%%%%%%%%%%%%%%%%%%%%%%%%%%
%%%%%%%%%%%%%%%%%%%%%%%%%%%%%%%%%%%%%%%%%%%

\end{document}